\documentstyle[aps,prl,multicol,floats,epsf]{revtex}

\begin{document}
\draft
\wideabs{

\title{Specific-heat evidence for strong electron correlations 
in the \\
thermoelectric material (Na,Ca)Co$_2$O$_4$}

\author{Yoichi Ando$^{1,2}$, N. Miyamoto$^{1,2}$, Kouji Segawa$^1$,
T. Kawata$^3$, and I. Terasaki$^3$}
\address{$^1$Central Research Institute of Electric Power
Industry, Komae, Tokyo 201-8511, Japan}
\address{$^2$Department of Physics, Science University of Tokyo,
Shinjuku-ku, Tokyo 162-8601, Japan}
\address{$^3$Department of Applied Physics, Waseda University,
Shinjuku-ku, Tokyo 169-8555, Japan}

\date{Received \today}
\maketitle

\begin{abstract}
The specific heat of (Na,Ca)Co$_2$O$_4$ is measured at low-temperatures
to determine the magnitude of the electronic specific-heat 
coefficient $\gamma$,
in an attempt to gain an insight into the origin of the unusually large thermoelectric power of this compound.
It is found that $\gamma$ is as large as $\sim$48 mJ/molK$^2$,
which is an order of magnitude larger than $\gamma$ of simple metals. 
This indicates that (Na,Ca)Co$_2$O$_4$ is a strongly-correlated electron 
system, where the strong correlation probably comes from the low-dimensionality and the frustrated spin structure.
We discuss how the large thermopower and its dependence on Ca doping 
can be understood with the strong electron correlations.
\end{abstract}

}

Recently, coexistence of a large thermopower 
($\sim$100 ${\rm \mu}$V/K at 300 K) and a low resistivity was 
found in a transition-metal oxide NaCo$_2$O$_4$ \cite{Terasaki}, 
which made this compound an attractive candidate for 
thermoelectric (TE) applications.
Normally, large thermopower is associated with materials with low carrier densities and the thermoelectric properties are optimized for systems
with typical carrier concentration of 10$^{19}$ cm$^{-3}$
\cite{Mahan}; 
on the other hand, 
NaCo$_2$O$_4$ has two-orders-of-magnitude larger carrier density 
($\sim$10$^{21}$ cm$^{-3}$) and yet
shows a thermopower comparable to that of the usual low-carrier-density
TE materials \cite{Terasaki}.  
The origin of the large thermopower in NaCo$_2$O$_4$ is
yet to be understood.

In NaCo$_2$O$_4$, Co ion has a mixed valence between 3+ and 4+.
Since NaCo$_2$O$_4$ is a layered system with a triangular lattice 
and Co$^{4+}$ has spins \cite{Terasaki}, 
it is expected that the interplay between charges and spins is playing 
a major role in producing the peculiar electronic properties of this
compound, as in the case of high-$T_c$ cuprates.
In those systems where Coulomb interactions or spin fluctuations are
important, it is often found that the electrons become
strongly correlated and thus the simple band picture is not well
applicable.
In fact, magnetotransport studies of NaCo$_2$O$_4$ found that the 
Hall coefficient has an opposite sign to the thermopower and is 
strongly temperature dependent \cite{Terasaki2}, which suggests the 
presence of a strong correlation in this system.
Therefore, to elucidate the origin of the large thermopower in 
NaCo$_2$O$_4$, it would be illuminating to determine the strength of the
electron correlations in NaCo$_2$O$_4$ by measuring the 
electronic specific heat.

In this paper, we report our specific-heat measurement of NaCo$_2$O$_4$
at low temperatures, which determines the
electronic specific-heat coefficient $\gamma$ of this system 
for the first time.
Since it has been reported that partial replacement of Na with Ca
systematically increases the thermopower \cite{Itoh}, we measured a series 
of (Na$_{1-x}$Ca$_x$)Co$_2$O$_4$ samples and investigated the change of 
$\gamma$ with Ca substitution.
Our results show that this system is indeed a strongly-correlated 
system with $\gamma$$\simeq$48 mJ/molK$^2$.
It is also found that the Ca substitution does not change the $\gamma$ 
value appreciably, while the Ca substitution reduces carrier 
concentration and increases the thermopower.
Based on these observations and by invoking a simple Drude picture,
we discuss that the large thermopower of NaCo$_2$O$_4$ is a result of
a large electronic specific heat.

\begin{figure}
\epsfxsize=0.9\columnwidth
\centerline{\epsffile{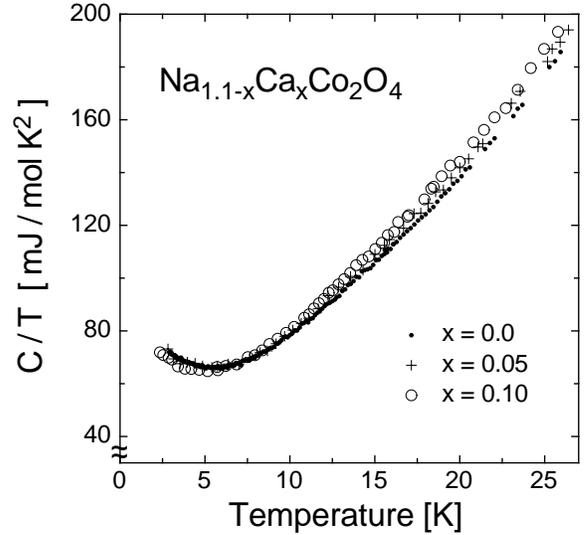}}
\vspace{0.2cm}
\caption{Plot of $C/T$ vs $T$ of the Na$_{1.1-x}$Ca$_x$Co$_2$O$_4$ samples 
with $x$=0.0, 0.05, and 0.10.}
\label{fig1}
\end{figure}

The samples used in this study are polycrystals prepared with a 
conventional solid-state reaction.
Starting powders of NaCO$_3$, CaCO$_3$, and Co$_3$O$_4$ are mixed
and calcined first at 860$^{\circ}$C for 12 hours, and then at 
800$^{\circ}$C for 6 hours.
Since it is known that Na tends to evaporate during the calcination,
which produces impurity phases in samples with (nominally) 
stoichiometric composition, we used samples with the composition of
Na$_{1.1-x}$Ca$_x$Co$_2$O$_4$ \cite{Itoh}.
The measurements were done on samples with three different Ca contents, 
$x$=0.0, 0.05, and 0.10.
The specific heat is measured using a standard quasi-adiabatic method
with a mechanical heat switch.
The mass of the samples used for the measurement is typically 1000 mg
and the heat capacity of the samples is always more than two orders of
magnitude larger than the addenda heat capacity.

Figure 1 shows the  specific heat $C$ of Na$_{1.1-x}$Ca$_x$Co$_2$O$_4$ 
in one decade of temperature range, from 2 K to 26 K.
One may immediately notice two features: 
(i) The magnitude of $C/T$, about 80 mJ/molK$^2$ at 10 K, 
is large compared to simple metals 
(for example, pure Cu has $C/T$ $\simeq$ 6 mJ/molK$^2$ at 10 K).
(ii) An unusual increase 
is observed at low temperatures for all $x$ values.

\begin{figure}
\epsfxsize=0.9\columnwidth
\centerline{\epsffile{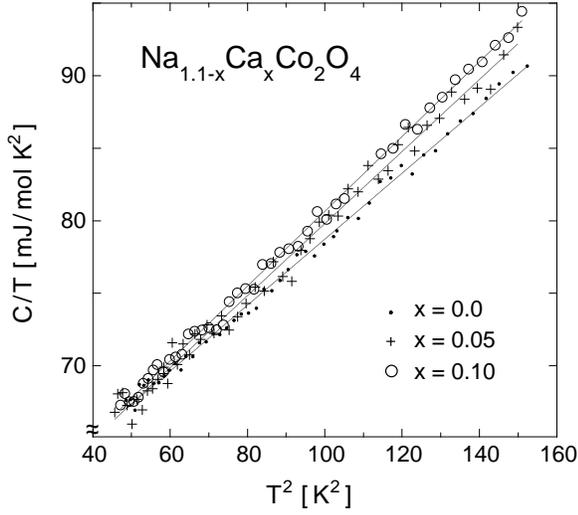}}
\vspace{0.2cm}
\caption{Plot of $C/T$ vs $T^2$ in the temperature range 7 - 12 K.
The solid lines are the straight-line fits to the data.}
\label{fig2}
\end{figure}

\begin{figure}
\epsfxsize=0.9\columnwidth
\centerline{\epsffile{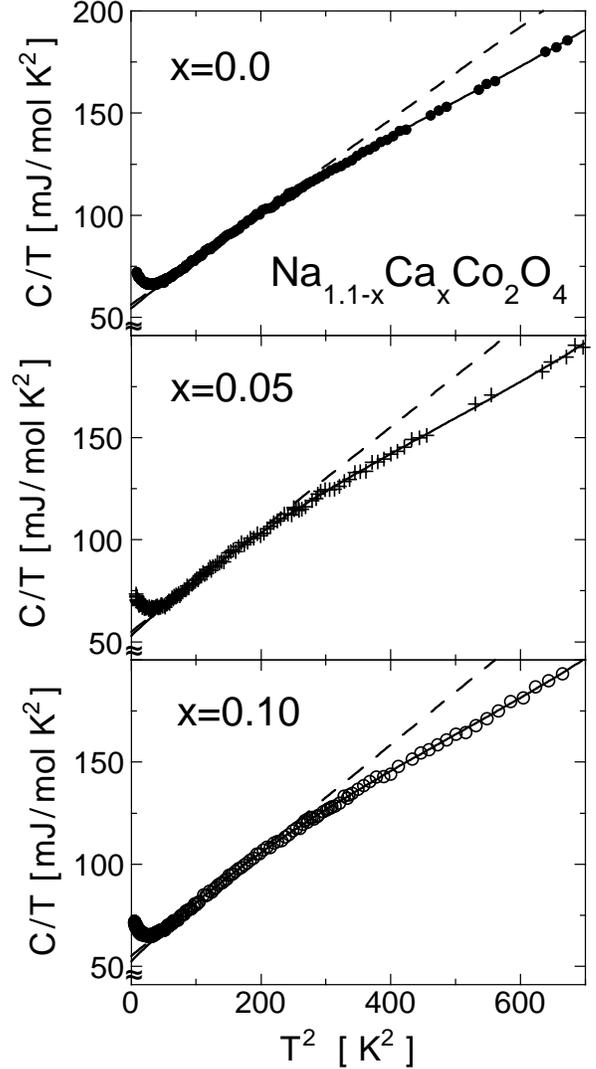}}
\vspace{0.2cm}
\caption{Plots of $C/T$ vs $T^2$ for $x$=0.0, 0.05, and 0.10. 
The data from 7 K to 26 K are fitted with Eq. (1) (solid lines). 
The straight-line fits to the data in a narrower range (7 - 12 K) 
are also shown (dashed lines).}
\label{fig3}
\end{figure}

As the first approximation, let us neglect the low-temperature increase 
in $C$ for the moment and analyze the data for $T>$7 K with the 
Debye formula.  
Since the temperature range to be analyzed is not
quite low enough, we should include higher-order terms for
the phonon specific heat and use the formula 
\begin{equation}
C/T = \gamma + \beta T^2  + \beta_5 T^4 + \beta_7 T^6.
\end{equation}
Figure 2 shows the result of the analysis which neglects
$\beta_5$ and $\beta_7$ (thereby the fit becomes a straight line 
in the plot of $C/T$ vs $T^2$) to show that the simple Debye formula 
without the higher-order lattice terms is moderately good to describe 
the data in the temperature region 7 - 12 K.
It is clear from Fig. 2 that $\gamma$ does not change appreciably
with $x$.
The values of $\gamma$ and $\beta$ obtained from the 
straight-line fits in Fig. 2 are listed in Table I.
The result of the analysis which uses the full formula of Eq. (1) is
shown in Fig. 3.
Clearly, Eq. (1) describes the data above 7 K very well and we obtained 
good fits in the temperature range 7 - 26 K for all three data sets.
The values of $\gamma$, $\beta$, $\beta_5$, and $\beta_7$ obtained from 
the fits in Fig. 3 are also listed in Table I.
The electronic specific-heat coefficient $\gamma$ obtained from this
analysis is relatively large, 52 - 54 mJ/molK$^2$, compared to simple 
metals where $\gamma$ is usually a few mJ/molK$^2$.
Interestingly, $\gamma$ does not change appreciably with 
$x$ within our range of resolution.

Perhaps as a better approximation, we next analyze our data using
the formula which includes the Schottky term:
\begin{eqnarray}
C/T &=& \gamma + \beta T^2 + \beta_5 T^4 + \beta_7 T^6 \nonumber\\
    & & + c_0 (T_0/T)^2  \frac{\exp (T_0/T)}{(\exp (T_0/T)+1)^2}   ,
\end{eqnarray}
where $T_0$ is the characteristic temperature for the
Schottky anomaly.
Figure 4 shows the result of the fit of the data to Eq. (2).
Apparently, the data in the whole temperature range measured 
(2 - 26 K) are well fitted with Eq. (2).
The fitting parameters are listed in Table II.
In both Tables I and II, the Debye temperatures calculated from 
$\beta$ are also listed.
Although the $\gamma$ values obtained with Eq. (2) tend to be smaller 
compared to the result of the simpler analysis using Eq. (1), 
the changes are only about 10\%. 
The values of $\gamma$ with this analysis are about 48 mJ/molK$^2$ and,
again, do not seem to be systematically correlated with $x$.
We note that the low temperature limit used for fitting the Schottky 
term is 2 K; extending the measurement to lower temperature is desirable 
for a better determination of both the Schottky anomaly and the 
$\gamma$ value.  It is possible that $\gamma$ becomes smaller than 
$\sim$48 mJ/molK$^2$ when the temperature range is extended.

\begin{figure}
\epsfxsize=0.9\columnwidth
\centerline{\epsffile{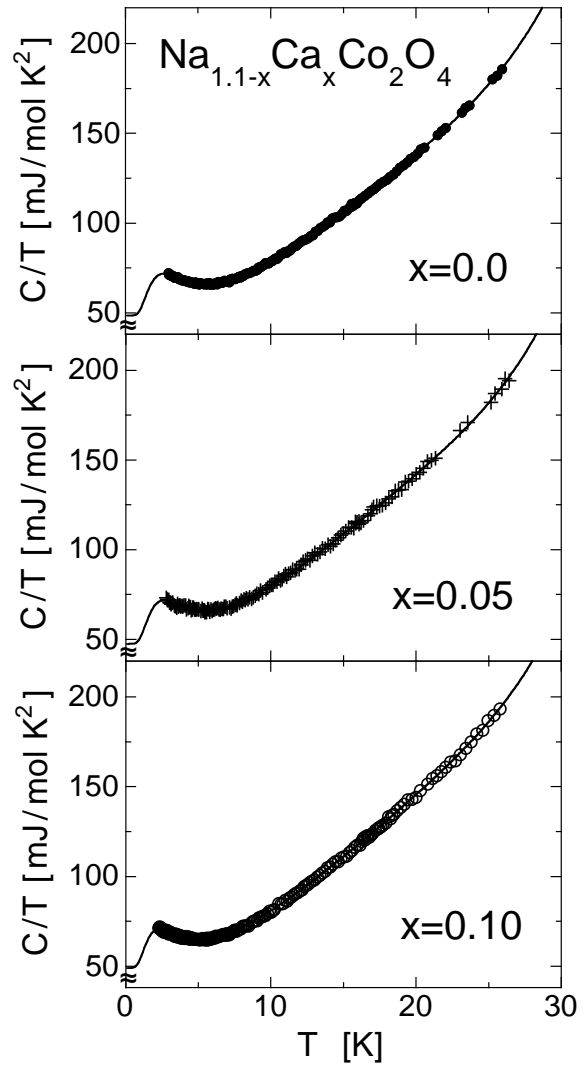}}
\vspace{0.2cm}
\caption{Plots of $C/T$ vs $T$ together with the fits (solid lines) 
to Eq. (2) which includes the Schottky term.}
\label{fig4}
\end{figure}

The above results indicate that the magnitude of the enhancement of
the density of states, which is represented in the magnitude of $\gamma$,
does not show a clear change with Ca substitution.
This is not a trivial result, because Ca substitution is expected to
reduce carrier density $n$ \cite{Itoh}.
As mentioned in the introduction, the origin of the strong correlation 
in this system is probably the frustration of the antiferromagnetically
interacting spins in the two-dimensional triangular lattice 
\cite{Terasaki}.
The magnetic susceptibility of NaCo$_2$O$_4$ shows a 
Curie-Weiss-like temperature
dependence \cite{Tanaka}, which suggests that spin fluctuations are 
actually appreciable in the magnetic properties.
Also, a negative magnetoresistance has been observed in the temperature
region where the resistivity does not show any localization behavior 
\cite{Terasaki2}, suggesting that the scattering from
spin fluctuations is playing a major role in the charge transport.  
If the spin fluctuations are indeed the source of the strong correlation,
one would expect the electron correlation to become stronger 
as the carrier concentration $n$ is decreased, 
because mobile carriers tend to destroy spin correlations.  
Since we expect Ca doping to reduce the free-electron density of 
states (DOS) through the reduction in $n$, 
the effect of increasing correlation and the decreasing
free-electron DOS upon Ca doping would tend to cancel with each other
in determining the electronic specific heat.
This might explain the apparent insensitivity of the observed
$\gamma$ to the Ca doping. 

Now let us briefly discuss the inference of our specific-heat result
to the large thermopower by employing the Drude picture for the
thermoelectric transport.
Although the simple Drude picture cannot explain all the aspects of the
complicated charge transport in NaCo$_2$O$_4$, it may help us 
to capture the basic physics for the enhancement of the thermopower.
The Seebeck coefficient $S$ of NaCo$_2$O$_4$ monotonically 
increases with $T$ \cite{Terasaki}, which suggests that the charge
transport is Fermi-liquid like and the Drude model is expected to be 
used as the first approximation (as opposed to many other 
strongly-correlated systems which show non-Fermi-liquid behavior).
In the simple Drude picture, the Seebeck coefficient $S$ is proportional 
to $c_e/n$, where $c_e$ is the electronic specific heat \cite{Ashcroft}.  
Thus, when the strong correlation enhances $c_e$, the Drude picture
predicts a large thermopower.  
More interestingly, the increase in $S$ of Na$_{1.1-x}$Ca$_x$Co$_2$O$_4$ 
with the Ca concentration $x$ is in qualitative agreement with the 
Drude picture, because our result shows  $\gamma$ (and thus $c_e$) 
to be almost unchanged upon Ca doping while $n$ decreases with 
increasing $x$.  This suggests that the simple Drude picture captures
the basic physics of the thermoelectric transport in NaCo$_2$O$_4$ and 
thus the large thermopower is actually a result of the 
strong electron correlation.
We note that an enhancement of the thermopower due to the large
effective mass has recently been discussed theoretically for 
strongly-correlated systems \cite{Palsson}.

It should be mentioned that the rather large change in $S$ with Ca doping
($S$ increases by about 20\% upon 0.1 of Ca substitution \cite{Itoh})
cannot be quantitatively explained by the simple Drude picture;
apparently, a more sophisticated model should be employed
for the full understanding of the large thermopower.
A semiclassical Boltzmann approach gives a formula for $S$,
which includes the energy dependence of the scattering time $\tau$,
$\frac{d\tau}{d\varepsilon}$.
\cite{Ashcroft2}.
Since the temperature dependence of $S$ changes with Ca doping
in Na$_{1.1-x}$Ca$_x$Co$_2$O$_4$ \cite{Itoh}, one can expect that 
$\frac{d\tau}{d\varepsilon}$ is actually changing with Ca doping, 
which introduces an additional factor in determining the Ca-doping 
dependence of $S$.

In summary, we found that the electronic specific-heat 
coefficient $\gamma$ of NaCo$_2$O$_4$ is
about 48 mJ/molK$^2$, which indicates that NaCo$_2$O$_4$ is a 
strongly-correlated system.  
No apparent correlation was found
between $\gamma$ and the $x$ value in 
Na$_{1.1-x}$Ca$_x$Co$_2$O$_4$, in which increasing $x$ reduces the 
carrier concentration $n$.  
The increase in the Seebeck coefficient $S$ with increasing $x$
and the apparent insensitivity of $\gamma$ to the change in $x$ 
together suggest that the simple Drude picture, which gives 
$S$$\sim$$c_e/n$, captures the basic physics for the enhancement 
of the thermopower, although quantitatively the simple Drude picture
is insufficient.
Therefore, it may be concluded that the larger thermopower of 
NaCo$_2$O$_4$ is a result of the strong electron correlation.

%

\newpage
\widetext

\begin{table}
\caption{Parameters obtained from the fit to Eq. (1) and the Debye temperature $\Theta_D$.}
\begin{tabular}{cccccc} 
$x$ & $\gamma$ (mJ/molK$^2$) & $\beta$ (mJ/molK$^4$) & 
$\beta_5$ (mJ/molK$^6$) & $\beta_7$ (mJ/molK$^8$) & $\Theta_D$ (K) \\
\tableline
0.0  & 56.0$\pm$0.2 & 0.227 &  &  & 393 \\
0.0  & 54.0$\pm$0.2 & 0.268 & $-1.95\times 10^{-4}$ & 
$1.31\times 10^{-7}$ & 372 \\
0.05 & 54.7$\pm$0.3 & 0.250 &  &  & 381 \\
0.05 & 52.8$\pm$0.4 & 0.296 & $-2.54\times 10^{-4}$ & 
$1.77\times 10^{-7}$ & 360 \\
0.10 & 54.8$\pm$0.2 & 0.259 &  &  & 376 \\
0.10 & 52.3$\pm$0.3 & 0.313 & $-2.75\times 10^{-4}$ & 
$1.86\times 10^{-7}$ & 353 \\
\end{tabular}
\end{table}

\begin{table}
\caption{Parameters obtained from the fit to Eq. (2) and the Debye
temperature $\Theta_D$.}
\begin{tabular}{cccccccc} 
$x$ & $\gamma$ (mJ/molK$^2$) & $\beta$ (mJ/molK$^4$) & 
$\beta_5$ (mJ/molK$^6$) & $\beta_7$ (mJ/molK$^8$) & 
$c_0$ (mJ/molK$^2$) & $T_0$ (K) & $\Theta_D$ (K) \\
\tableline
0.0  & 48.4$\pm$0.3 & 0.311 & -3.06$\times 10^{-4}$ & 2.19$\times 10^{-7}$ 
& 16.9 & 8.14 & 354 \\
0.05 & 47.5$\pm$0.8 & 0.334 & -3.48$\times 10^{-4}$ & 2.49$\times 10^{-7}$ 
& 17.4 & 8.13 & 346 \\
0.10 & 49.0$\pm$0.3 & 0.336 & -3.29$\times 10^{-4}$ & 2.27$\times 10^{-7}$ 
& 14.0 & 7.24 & 345 \\
\end{tabular}
\end{table}

\end{document}